\documentclass[english]{JHEP3}

\usepackage{amsmath,amssymb,bbm}

\usepackage{graphicx}
\usepackage{axodraw}

\newcommand{\Tr}{\mathrm{Tr}}

\newcommand{\E}{\mathrm{e}}

\newcommand{\be}{\begin{eqnarray}}
\newcommand{\ee}{\end{eqnarray}}

\newcommand{\Tc}{T_{\text{cr}}}
\newcommand{\yb}{\bar\psi}
\newcommand{\xsb}{$\chi$SB}
\newcommand{\Nf}{N_{\text{f}}}
\newcommand{\Nc}{N_{\text{c}}}
\newcommand{\pat}{\partial_t}
\newcommand{\Eqref}[1]{Eq.~\eqref{#1}}

\newcommand{\LQCD}{\Lambda_{\text{QCD}}}

\newcommand{\fslash}{\hspace{-0.1ex} \slash }
\def\slash#1{\setbox0=\hbox{$#1$}               
   \dimen0=\wd0                                 
   \setbox1=\hbox{/} \dimen1=\wd1               
   \ifdim\dimen0>\dimen1                        
      \rlap{\hbox to \dimen0{\hfil/\hfil}}      
      #1                                        
   \else                            
      \rlap{\hbox to \dimen1{\hfil$#1$\hfil}}   
      /                                         
   \fi}                                         %

\title{Scaling laws near the conformal window of many-flavor QCD}

\author{Jens~Braun and Holger~Gies\\
Theoretisch-Physikalisches Institut, Friedrich-Schiller-Universit\"at Jena,\\
Max-Wien-Platz 1, D-07743 Jena, Germany}

\abstract{
We derive universal scaling laws for physical observables such as the critical
temperature, the chiral condensate, and the pion decay constant as a function
of the flavor number near the conformal window of many-flavor QCD in the
chiral limit. We argue on
general grounds that the associated critical exponents are all interrelated
and can be determined from the critical exponent of the running gauge coupling at
the Caswell-Banks-Zaks infrared fixed point. We illustrate our findings with the aid
of nonperturbative functional Renormalization Group (RG) calculations and low-energy QCD models.}

\begin{document}


%
\section{Introduction}

Many-flavor QCD has recently attracted a great deal of attention for a number
of reasons: first, as a controllable deformation of real QCD, it can teach
important lessons about the chiral structure of QCD-like theories. Second, it
serves as a building block for alternative technicolor-like scenarios for the
Higgs sector. And third, it exhibits a quantum phase transition from the
QCD-like chirally broken to a conformal phase as a function of the flavor
number and thus gives rise to interesting quantum critical behavior. 

Adding $\Nf$ massless quark flavors to a nonabelian SU($\Nc$) gauge theory
increases the screening property of fermionic fluctuations. An obvious
consequence is the loss of asymptotic freedom for
$\Nf>\Nf^{\text{a.f.}}:=\frac{11}{2} \Nc$ ($=16.5$ for SU(3)). Already at
smaller $\Nf$, $\Nf>\frac{34\Nc^3}{13\Nc^2-3}$ ($\simeq 8.05$ for SU(3)), the second $\beta$ function
coefficient changes sign inducing an infrared (IR) attractive fixed point of
the gauge coupling $\alpha_\ast>0$ \cite{Caswell:1974gg}. For small
$\Nf^{\text{a.f.}}- \Nf>0$, the fixed-point value is small. This gives access to a
perturbative analysis, suggesting that the system approaches a conformally
invariant limit in the deep IR \cite{Banks:1981nn}.
For smaller $\Nf$ (such as real QCD), this Caswell-Banks-Zaks fixed point is
destabilized due to the spontaneous break-down of chiral symmetry, resulting
in massive fermionic excitations, strongly-coupled glue and massless Goldstone
bosons. 

The above considerations propose the existence of a critical flavor number
$\Nf^{\text{cr}}$, separating the chiral-symmetry-broken phase from the
conformal phase. Theories with a flavor number $\Nf$ satisfying
$\Nf^{\text{cr}}\leq \Nf<\Nf^{\text{a.f.}}$ are said to be in the conformal
window. The flavor number $\Nf$ therefore serves as a control parameter for a
quantum phase transition.  Investigations of this phase structure have been
performed by continuum methods
\cite{Miransky:1996pd,Appelquist:1996dq,Appelquist:1997dc,Schafer:1996wv,Velkovsky:1997fe,Appelquist:1998rb,%
  Harada:2000kb,Sannino:1999qe,Harada:2003dc,Gies:2005as,Braun:2005uj,Braun:2006jd,%
  Poppitz:2009uq,Armoni:2009jn,Sannino:2009qc}, as well as lattice simulations
\cite{Kogut:1982fn,Gavai:1985wi,Fukugita:1987mb,Brown:1992fz,Damgaard:1997ut,%
  Iwasaki:2003de,Catterall:2007yx,Appelquist:2007hu,Deuzeman:2008sc,Deuzeman:2009mh,Appelquist:2009ty,%
  Fodor:2009wk,Fodor:2009ff,Pallante:2009hu}. Recent results have collected a substantial body
of evidence for the existence of the conformal phase and a critical flavor
number, defining the onset of the conformal phase: lattice simulations have
provided evidence that $8<\Nf^{\text{cr}}\leq 12$
\cite{Appelquist:2007hu,Deuzeman:2008sc,Deuzeman:2009mh,Appelquist:2009ty,Fodor:2009wk},
even though the case $\Nf=12$ is controversial, see \cite{Fodor:2009ff}.  On
the present level of accuracy, these results go well together with an earlier
quantitative estimate from a combination of four-loop perturbation theory with
the functional Renormalization Group (RG) which yields $\Nf^{\text{cr}}=10.0
\genfrac{}{}{0pt}{}{+1.6}{-0.7}$ \cite{Gies:2005as}.

Beyond the precise location of this quantum critical point on the $\Nf$ axis,
the critical behavior in the vicinity of the fixed point is expected to show
several peculiarities: for $\Nf<\Nf^{\text{cr}}$ in the chirally broken phase,
a standard chiral phase transition can be anticipated to occur at finite
temperature. The critical phenomena near this finite-$T$ chiral phase
transition are expected to be determined by the precise symmetries of the
order parameter, i.e., the chiral condensate, defining the universality class
of the transition (e.g., SU(2)$_{\text{L}}\otimes\,$SU(2)$_{\text{R}}\simeq$O(4)
for $\Nf=2$) \cite{Pisarski:1983ms}. Near the critical temperature an effective description in terms of 
a Ginzburg-Landau-type effective potential is expected to hold even on a 
quantitative level. By contrast, the zero-temperature quantum phase transition 
as a function of the control parameter $\Nf$ may not have a continuous Ginzburg-Landau
description. In particular, there appear to be no light scalar states in terms
of which an effective theory could be constructed on the conformal side
$\Nf\gtrsim\Nf^{\text{cr}}$ of the phase transition
\cite{Miransky:1996pd,Appelquist:1996dq,Chivukula:1996kg}, even though the
order parameter (chiral condensate) should change continuously across the
phase transition. 

Plotting the chiral-phase-transition temperature versus the
control parameter $\Nf$ yields the phase boundary in the $(T,\Nf)$ plane which
has first been explored in \cite{Braun:2005uj,Braun:2006jd} using the
functional RG. Near the critical flavor number, an intriguing relation between
the shape of the phase boundary and RG properties of the Caswell-Banks-Zaks
fixed point has been identified. This relation connects the scaling of the
critical temperature with the critical exponent of the gauge coupling near the
Caswell-Banks-Zaks fixed point. In the present work, we argue that this
scaling relation can be extended to further physical observables such as the
chiral condensate or the pion decay constant. As these observables are
accessible to a variety of nonperturbative methods, our results suggest that
the corresponding scaling can become a useful tool to study the phase
transition to the conformal phase quantitatively. 

Our main arguments are based on very general considerations and involve only
few assumptions about the RG structure of the theory. These are illustrated in
Sect.~\ref{sec:fewflavor} for the simple few-flavor case and in
Sect.~\ref{sec:manyflavor} for many flavors near the conformal window. These
arguments are then made more concrete with the aid of functional RG
calculations in a derivative expansion of the QCD effective action, or simply
within low-energy QCD models in Sect.~\ref{sec:funRGresults}. Throughout this
work, we concentrate on the chiral phase transition even though we also expect
an impact of the confining nature of the theory on the properties of the
system near criticality. However, as we work in the chiral limit, there is no
good order parameter for confinement, implying that nonanalyticities in the
correlations are rather dominated by the chiral degrees of freedom.

\section{Few-flavor QCD and the role of scale fixing}
\label{sec:fewflavor}

QCD in the chiral limit of zero quark masses depends only on one parameter in
the Euclidean Lagrangian, namely the gauge coupling $g$,
\begin{equation}
\mathcal{L}= \frac{1}{4g^2 } F_{\mu\nu}^a F_{\mu\nu}^a + i \bar\psi \gamma_\mu
D_\mu[A] \psi.
\label{eq:LQCD}
\end{equation}
In the quantum theory, the gauge coupling has to be fixed at a certain
momentum scale in terms of a renormalization condition. The renormalization
group finally trades the gauge coupling fixed at an arbitrary scale for one
single parameter $\LQCD$ of mass dimension one which sets the mass scale for
all physical observables of the theory. In other words, all physical
observables respond trivially to a variation of $\LQCD$ according to their
engineering dimension. In units of $\LQCD$, the theory is completely fixed. 

In order to discuss the dependence on quantities such as the flavor number, it
is important to emphasize that a variation of the flavor number does not
merely correspond to a change of a parameter of the theory. It rather
corresponds to changing the theory itself. In particular, there is no unique
way to unambiguously compare theories of different flavor number with each
other, as different theories may have different scales $\LQCD$. 

For instance, it might seem natural to compare theories with different flavor
numbers at fixed $\LQCD$ with each other. But $\LQCD$ itself is not a
direct observable, so that such a comparison is generically inflicted with
theoretical uncertainties. Moreover, $\LQCD$ is regularization-scheme
dependent which can affect comparisons between different theoretical methods,
say, lattice and continuum results. Another option could be a scale fixing in
the deep perturbative region, say, at the $Z$ mass pole by fixing
$\alpha(M_Z)$. However, theories with different flavor numbers then exhibit a
different perturbative running, such that IR observables vary because of both
high-scale perturbative as well as non-perturbative evolution. 

Instead, we propose to choose a mid-momentum scale for the scale fixing, as
the high-scale perturbative running is then separated from the more
interesting non-perturbative dynamics. In this work, we fix the theories at
any $\Nf$ by keeping the running coupling at the $\tau$ mass scale fixed to
$\alpha(m_\tau)=0.322$. Even though also this choice is scheme dependent,
these dependencies should be subdominant, as they follow a perturbative
ordering. In general, fixing the scale via the coupling is a prescription
which is well accessible by many nonperturbative methods. 

Let us now present a simple argument that illustrates how $\Nf$ dependencies
of physical observables can be understood in the limit of small $\Nf$. As
already stated above, all IR observables such as the critical temperature
$\Tc$, pion decay constant $f_\pi$,  chiral condensate $\langle
\bar\psi\psi\rangle^{1/3}$, and model-dependent concepts such as the
constituent quark mass, are proportional to $\LQCD$. The latter on the one
hand can be read off from the UV behavior of the running coupling, $\alpha(k)
\sim 1/\ln (k/\LQCD)$ for large $k$. On the other hand, the value of $\LQCD$
can be associated with the position of the Landau pole in perturbation
theory.\footnote{Of course, this statement has to be taken with care, since
  $\LQCD$ is a meaningful scale, whereas the Landau pole is simply an artifact
  of perturbation theory.} In this simple sense, the artificial Landau pole
can be taken as an estimate for the scaling of physical observables. In
one-loop RG-improved perturbation theory, the position of the Landau pole can
be read off from 
\begin{eqnarray}
0 &\leftarrow &\frac{1}{\alpha(\LQCD)}=\frac{1}{\alpha(\mu_0)} +4\pi b_0\, \ln
\frac{\LQCD}{\mu_0},\nonumber\\
&& b_0=\frac{1}{8\pi^2} \left( \frac{11}{3} \Nc
-\frac{2}{3} \Nf \right), \label{eq:LandauPole}
\end{eqnarray}
where $\mu_0$ denotes a perturbative scale, such as
$m_\tau,M_Z,\dots$. Solving this equation for $\LQCD$ and expanding the result
for small $\Nf$ leads us to
\begin{eqnarray}
  \LQCD &\simeq&\mu_0\, \E^{-\frac{1}{4\pi b_0
      \alpha(\mu_0)}} \nonumber\\
  &\simeq&\mu_0\, \E^{-\frac{6\pi}{11\Nc \alpha(\mu_0)}} \left(
  1-\epsilon \Nf + \mathcal O ((\epsilon \Nf)^2)\right). \nonumber
\end{eqnarray}
Choosing $\mu_0=m_\tau$, we find $\epsilon = \frac{12 \pi}{121 \Nc^2
  \alpha(\mu_0)} \simeq 0.107$ for $\Nc=3$. Two conclusions can immediately be
drawn: first, $\LQCD$ can be expanded in $\Nf$ and has a generically
nonvanishing linear term; second, for the present way of scale fixing, the
linear behavior should be a reasonable approximation for finite values of
$\Nf$, say $\Nf\lesssim 4$, as the expansion parameter $\epsilon$ is small.

As $\LQCD$ sets the scale for all dimensionful IR observables, we are
tempted to conclude that all IR observables scale linearly with $\Nf$ for
small $\Nf$ with the same proportionality constant $\epsilon$. This is, of
course, a bit too simplistic, as the dynamics which establishes the value of
the IR observables generically carries an $\Nf$ dependence as well. E.g., the
chiral symmetry-breaking dynamics depends on the number of light
mesonic degrees of freedom, which is an $\Nf$-dependent quantity. Detailed
quantitative model studies \cite{Braun:2009si}, however, demonstrate that the scaling
of the critical temperature does not receive strong corrections of this type
and indeed scales according to
\begin{equation}
\Tc= T_0(1- \epsilon \Nf + \mathcal O((\epsilon \Nf)^2)),
\label{eq:lowNfTc}
\end{equation}
where $T_0$ is a dimensionful proportionality constant. We conclude that the
phase boundary in the $(T,\Nf)$ plane has a linear shape for small $\Nf$ which
can mainly be understood as a result of the perturbative $\Nf$ scaling of
$\LQCD$. Note that this observation is consistent with lattice simulations~\cite{Karsch:2000kv} 
and has been exploited for parameter fixing in PNJL/PQM-model studies~\cite{Schaefer:2007pw}. In the following, 
we will show that the shape of the phase boundary as well as the $\Nf$ scaling of other observables 
for large $\Nf$ can also be understood from simple scaling arguments which this time 
follow from general properties of the nonperturbative domain. 
\section{Scaling in many-flavor QCD near the conformal window}
\label{sec:manyflavor}
In this section, we review, detail and extend the scaling arguments presented
in \cite{Braun:2005uj,Braun:2006jd}, leading to universal relations near the
conformal window. Whereas the $\Nf$ scaling in the few-flavor case essentially
follows from analyticity of the observables in $\Nf$ also for $\Nf$ near zero,
the $\Nf$ dependence near the conformal window is clearly nonperturbative in
$\Nf$.  The mere existence of the conformal window requires a sufficient
amount of fermionic screening, i.e., a sufficient amount of fermionic degrees
of freedom $\Nf\geq \Nf^{\text{cr}}$.

The lower end of the conformal window is characterized by the onset of chiral
symmetry breaking. Whereas the coupling approaches the Caswell-Banks-Zaks
fixed point $g_\ast^2$ in the conformal window, chiral symmetry breaking
destabilizes this fixed point below the critical flavor number
$\Nf^{\text{cr}}$. This suggests the existence of a critical coupling
$g_{\text{cr}}^2$. If $g^2>g_{\text{cr}}^2$ at some scale, the system will be
triggered to run into the chirally broken regime. 

For a monotonous coupling flow, the value of the Caswell-Banks-Zaks fixed
point $g_\ast^2$ or its nonperturbative variant corresponds to the maximum
possible coupling strength of the system in the conformal window.\footnote{
  The following scenario does not apply if the running coupling overshoots,
  develops a local maximum and approaches the fixed point from above which
  requires a double-valued $\beta$ function.  This behavior is seen, e.g., in
  MOM-scheme running couplings derived from the ghost-gluon vertex using
  truncated Dyson-Schwinger equations
  \cite{Alkofer:2000wg} for pure Yang-Mills
  theory. If such a scenario held in many-flavor QCD, the occurrence of \xsb,
  i.e., whether or not the fermion sector becomes critical for a given $\Nf$,
  would depend quantitatively on the details of the coupling flow, making it
  difficult to extract universal features.  However, we do not expect such a
  running to occur in many-flavor QCD near the conformal window as the fast
  running of fermions in the near-critical region reduces the fermionic
  screening contributions, thus supporting a monotonous increase of the gauge
  coupling. Also, such a behavior is not observed near the upper end of the
  conformal window where perturbation theory is expected to hold.  } As both
$g_\ast^2$ and $g_{\text{cr}}^2$ depend on the number of flavors, the
condition $g_\ast^2(\Nf^{\text{cr}})=g_{\text{cr}}^2(\Nf^{\text{cr}})$ defines
the lower end of the conformal window and thus the critical flavor number, see
left panel of Fig.~\ref{fig:sketches} for an illustration.
%
\FIGURE[t]{
\includegraphics[width=0.47\linewidth]{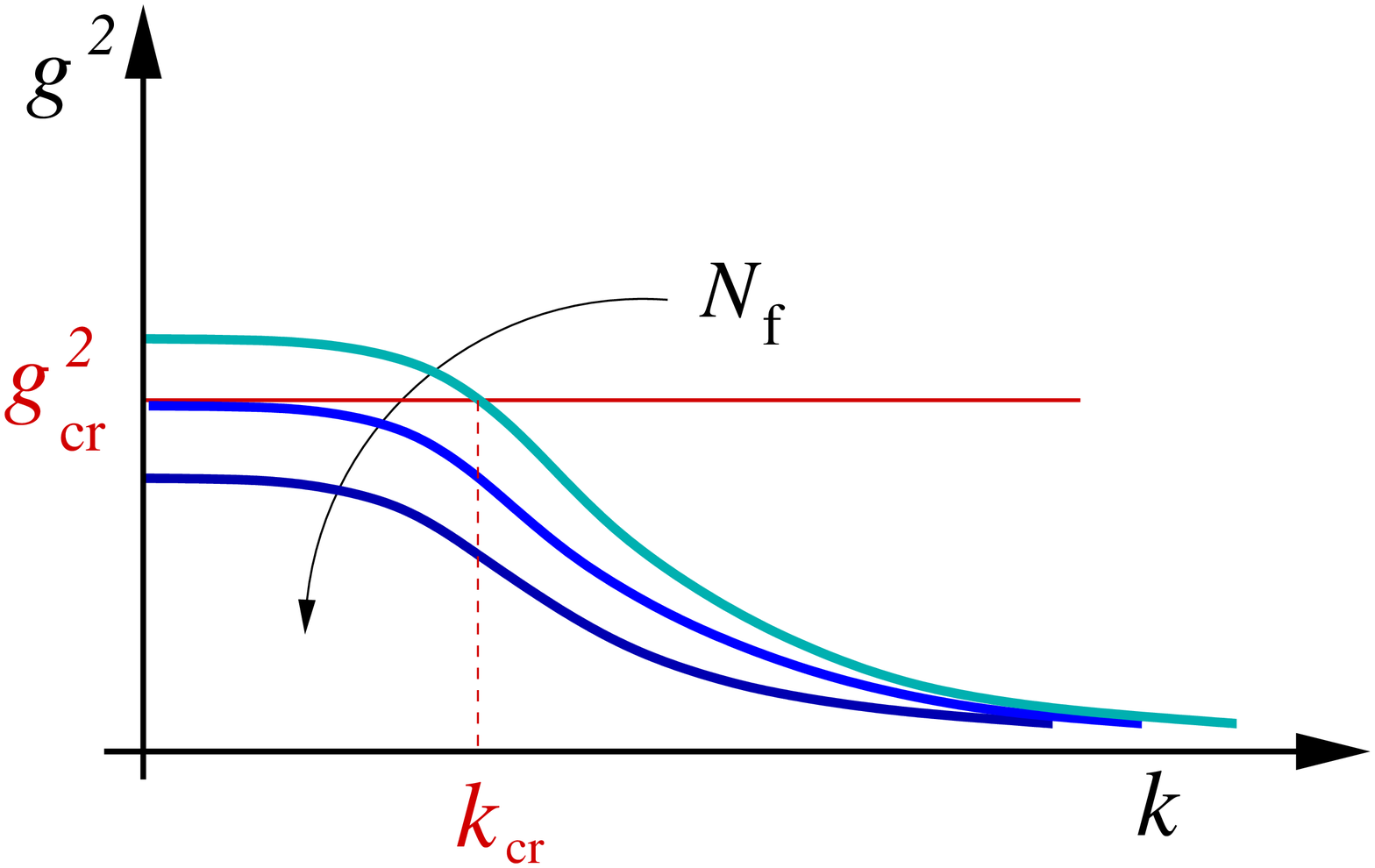}
  \hspace*{.5cm}
\includegraphics[width=0.41\linewidth,height=0.3\linewidth]{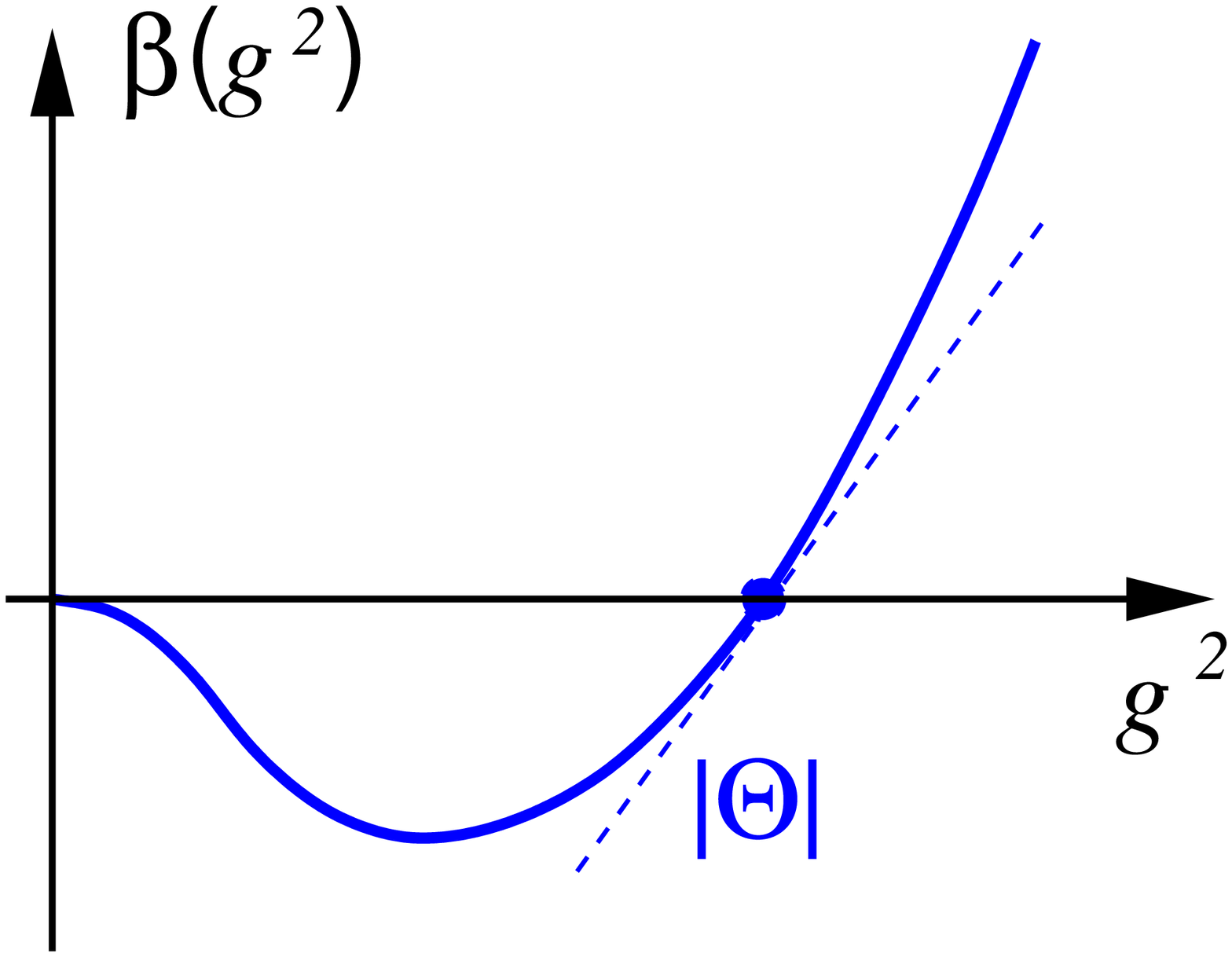}\quad
\caption{Left panel: illustration of the IR running of the gauge coupling in
  comparison with the critical coupling $g_{\text{cr}}^2$ for \xsb. Below the
  conformal window, $\Nf<\Nf^{\text{cr}}$, $g^2$ exceeds the critical value, triggering the approach
  to \xsb. For increasing flavor number, the IR fixed-point value $g_\ast^2$
  drops below the critical value denoting the onset of the conformal
  window. Right panel: sketch of the $\beta$ function of the running
  coupling. The slope of the $\beta$ function at the IR fixed point
  corresponds to minus the critical exponent $\Theta$, cf. Eq.~(\ref{eq:FPR}).
  }
\label{fig:sketches}
}
%
For $g_\ast^2>g_{\text{cr}}^2$, the model is below the conformal window and
runs into the broken phase. Slightly below the conformal window, the running
coupling $g^2$ exceeds the critical value while it is in the attractive domain
of the IR fixed point $g_\ast^2$. The flow in this fixed-point regime can approximately 
be described by the linearized $\beta$ function
\begin{equation}
\beta_{g^2}\equiv \pat g^2 =-\Theta\, (g^2-g_\ast^2)+ \mathcal
O((g^2-g_\ast^2)^2), \label{eq:FPR}
\end{equation}
where $t=\ln(k/\Lambda)$ with $k$ being a suitable RG scale and $\Lambda$ defining
the UV cutoff. The universal ``critical exponent'' $\Theta$ denotes (minus) the first
expansion coefficient. We know that $\Theta<0$, since the fixed point is IR
attractive, see right panel of Fig.~\ref{fig:sketches}. In general, the critical exponent depends 
on $\Nf$, $\Theta=\Theta(\Nf)$.  The solution to Eq.~\eqref{eq:FPR} for the running coupling in 
the fixed-point regime reads
\begin{equation}
g^2(k)=g_\ast^2-\left(\frac{k}{k_0}\right)^{-\Theta}, \label{eq:FPsol}
\end{equation}
where the scale $k_0$ is implicitly defined by a suitable initial condition
and is kept fixed in the following. It corresponds to a scale where the system
is already in the fixed-point regime, and otherwise plays the same role as the
renormalization scale $\mu_0$ in Eq.~\eqref{eq:LandauPole}; in particular, physical observables are
independent of $k_0$. For the present fixed-point considerations, it provides
for all dimensionful scales in the following. But knowing the full RG
trajectory, $k_0$ is related to $\mu_0$ and thus, say,  to the initial
$\tau$ mass scale by RG evolution.

Our criterion for \xsb\ is that $g^2(k)$ should exceed $g_{\text{cr}}^2$ for
some value of $k\leq k_{\text{cr}}$. From Eq.~\eqref{eq:FPsol} and the condition
$g^2(k_{\text{cr}})= g^2_{\text{cr}}$, we derive the estimate valid in the
fixed-point regime
\begin{equation}
k_{\text{cr}}\simeq k_0\, (g_\ast^2
-g_{\text{cr}}^2)^{-\frac{1}{\Theta}}. \label{eq:kcrest}
\end{equation}
This scale $k_{\text{cr}}$ now takes over the role of the fixed renormalization
scale $\mu_0=m_\tau$ in the small-$\Nf$ argument given above: it sets the
scale for the critical temperature $T_{\text{cr}}\sim k_{\text{cr}}$ with a
proportionality coefficient provided by the solution of the full flow. The
last step of the argument goes along with the estimate that the IR fixed-point
value $g_\ast^2$ roughly depends linearly on $\Nf$. More precisely, we assume
that the $\Nf$ dependence of the coupling quantities can be linearized near
the critical flavor number.  From Eq.~\eqref{eq:kcrest}, we thus find the
relation
\begin{equation}
T_{\text{cr}}\sim k_0 |\Nf -\Nf^{\text{cr}}|^{-\frac{1}{\Theta}},
\label{eq:TcrTheta}
\end{equation}
which is expected to hold near $\Nf^{\text{cr}}$ for
$\Nf\leq\Nf^{\text{cr}}$. Here, $\Theta$ should be evaluated at
$\Nf^{\text{cr}}$.\footnote{Accounting for the $\Nf$ dependence of $\Theta$ by
  an expansion around $\Nf^{\text{cr}}$ yields mild logarithmic corrections to
  Eq.~\eqref{eq:TcrTheta}.}  Relation \eqref{eq:TcrTheta} is an analytic
prediction for the shape of the chiral phase boundary in the ($T,\Nf$) plane
of QCD. This result is remarkable for a number of reasons: first, it relates two
universal quantities with each other: the phase boundary and the IR critical
exponent. Second, it establishes a quantitative connection between the chiral
structure ($T_{\text{cr}}$) and the IR gauge dynamics ($\Theta$). Third, it is
a parameter-free prediction following essentially from scaling arguments. 

As Eq.~\eqref{eq:TcrTheta} relates two universal quantities, it is important to
understand to what extent the underlying argument makes use of non-universal
but scheme-dependent quantities. Of course, the running coupling is a strongly
scheme-dependent concept, and so is the value of the IR fixed
point $g_\ast^2$. However, the existence of the fixed point as well as the
value of the critical exponent are scheme independent.\footnote{The running coupling in
  addition is definition dependent. Here, we assume that the running coupling
  used in the discussion provides for a reasonable measure of the interaction
  between the gauge and the quark sector which manifestly exhibits the IR
  fixed point in the conformal window.} Also, the value of the critical
coupling $g_{\text{cr}}^2$ is scheme dependent. Nevertheless this scheme dependence has
to cancel against that of the running coupling itself, as whether or not
\xsb\ occurs is a universal feature of the system.

Furthermore, we have implicitly neglected all dependencies of the running on
scales other than the RG scale $k$. At finite temperature, there will, of
course, be dependencies of the coupling on $T$ in particular at low scales
where $T/k$ becomes large. The relevant scale for the above argument, however,
is the scale $k_{\text{cr}}$ where \xsb\ is triggered. We expect that this
scale is generically somewhat larger than the temperature as long as we are in
the \xsb\ phase or approach the phase transition from below. Indeed, this
assumption turns out to hold in all model calculations as well as in the RG
results described below.  This allows us to ignore the $T$ dependence of the
running coupling $g^2$ and of the critical coupling $g_{\text{cr}}$.

Let us now generalize these considerations to other physical observables such
as the chiral condensate or the pion decay constant. To be more specific, we
are interested in the scaling behavior of physical observables as a function
of the number of massless quark flavors $\Nf$ at vanishing temperature. In
this case, the above argument can be followed straightforwardly, where the
determination of the critical (RG) scale $k_{\text{cr}}$ in
Eq.~\eqref{eq:kcrest} plays a prominent role. In terms of low-energy effective
theories, the scale $k_{\text{cr}}$ can be viewed as an ultraviolet (UV)
cutoff. From Eqs.~\eqref{eq:kcrest} and~\eqref{eq:TcrTheta}, it follows
immediately that the critical scale and therewith this UV cutoff
$\Lambda_{\text{eff}}$ of the low-energy sector is tightly related to the
critical flavor number:
\begin{equation}
\Lambda_{\text{eff}}\sim 
k_{\text{cr}}\simeq k_0\, |\Nf -\Nf^{\text{cr}}|^{-\frac{1}{\Theta}} \label{eq:kcrest2}.
\end{equation}
On the other hand, observables ${\mathcal O}$ with mass dimension $d_{\mathcal
  O}$ which are computable in this effective field theory defined by the
fixed-point regime are necessarily related to the UV cutoff
$\Lambda_{\text{eff}}$ in a simple manner,
\begin{equation}
{\mathcal O} \simeq c_{\mathcal O} \Lambda_{\text{eff}}^{d_{\mathcal O}}\,,\label{eq:obs_general}
\end{equation}
where  $c_{\mathcal O}$ is a numerical constant which depends on the details of
the theory, e.g., the number of colors $\Nc$ and also the number of flavors
$\Nf$. Here, we have assumed that a general separation of scales holds in the
sense that all effective UV parameters of the low-energy effective theory are
fully determined by the quark-gluon dynamics in the mid- and high-momentum
regime. Combining Eqs.~\eqref{eq:kcrest2} and~\eqref{eq:obs_general}, we find
\begin{equation}
{\mathcal O} \simeq k_0^{d_{\mathcal O}} |\Nf -\Nf^{\text{cr}}|^{-\frac{d_{\mathcal O}}{\Theta}}.
\label{eq:observ_scaling}
\end{equation}
This relation extends the scaling properties of the critical temperature near
the conformal window found above in \Eqref{eq:TcrTheta} to that of other
physical IR observables. Again the universal scaling of these observables as a
function of $\Nf$ is related to the critical exponent $\Theta$ of the running
coupling.

\section{Functional RG results}
\label{sec:funRGresults}

The critical temperature as a function of $\Nf$ has first been computed in
\cite{Braun:2005uj,Braun:2006jd} in the framework of the functional RG
\cite{Wetterich:1992yh} (see \cite{Reuter:1996ub} for reviews on the
functional RG in gauge theories). In this section, we briefly review these
results and discuss them in the light of the scaling relations. Result on further IR
observables will be discussed in the next section.

In \cite{Braun:2005uj,Braun:2006jd}, the RG flow of QCD starting from the
microscopic degrees of freedom in terms of quarks and gluons was studied
within a covariant derivative expansion, approaching the critical
\xsb\ temperature from above. A crucial ingredient for \xsb\ are the
scale-dependent gluon-induced quark self-interactions of the type
\begin{equation}
\Gamma_{\psi,\text{int}}=\int \hat\lambda_{\alpha\beta\gamma\delta}
\yb_\alpha \psi_\beta \yb_\gamma \psi_\delta, \label{psitrunc}
\end{equation}
where $\alpha,\beta,\dots$ denote collective indices including color, flavor,
and Dirac structures. These four-fermion interactions are set to zero at the
initial UV scale, $\hat\lambda_{\alpha\beta\gamma\delta}|_{k\to\Lambda}\to
0$. This guarantees that the $\hat\lambda$'s at $k<\Lambda$ are solely
generated by quark-gluon dynamics from first principles (e.g., by 1PI ``box''
diagrams with 2-gluon exchange).  This is an important difference to models
such as the Nambu-Jona-Lasinio (NJL) model, where the $\hat\lambda$'s are
independent input parameters.

As we approach the chiral phase transition temperature from above, the
derivative expansion with local ``point-like'' interactions is a
self-consistent approximation. This corresponds to replacing the momentum
structure of the $\hat\lambda$'s by the overall $k$ dependence on the RG
scale. This approximation has been successfully tested quantitatively in
\cite{Gies:2005as} by verifying the resulting insensitivity of the many-flavor
quantum phase transition on the momentum regularization. In the chirally
broken regime, this approximation breaks down as, e.g., mesons manifest
themselves as momentum singularities in these vertices.
These restrictions result in a total number of four linearly independent
couplings $\hat\lambda_i$ \cite{Gies:2003dp}.  Introducing the dimensionless couplings
$\lambda_i=k^2 \hat\lambda_i$, the corresponding $\beta$ functions read
\begin{equation}
\pat\lambda_i=2\lambda_i -b_{ij} \lambda_j g^2 - A_{ijk}\lambda_j \lambda_k -
c_i g^4,
\label{eqlambda}
\end{equation}
where the coefficients $A$, $b$, $c$ depend on the temperature, number of
quark flavors $N_f$ and number of colors $N_c$; for explicit representations,
see \cite{Gies:2005as,Braun:2005uj,Braun:2006jd}.

\begin{figure}[!t]
\begin{center}
\scalebox{0.8}[0.8]{
\begin{picture}(190,160)(80,40)
\includegraphics[width=12.5cm,height=7cm]{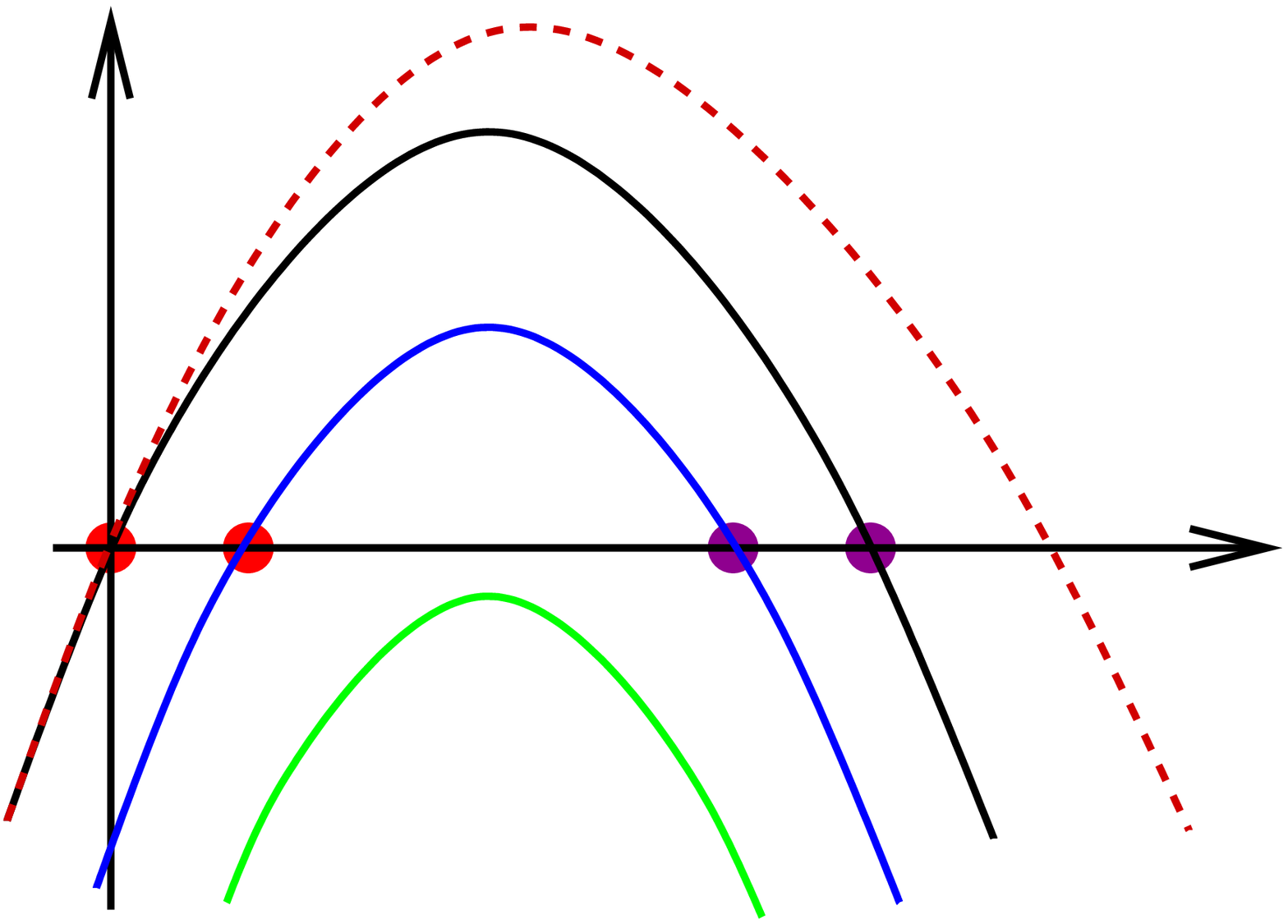}
\Text(-10,+60)[c]{\scalebox{1.6}[1.6]{$\lambda_i$}}
\Text(-345,160)[c]{\scalebox{1.6}[1.6]{$\pat\lambda_i$}}
\Text(-220,155)[c]{\scalebox{1.6}[1.6]{$g=0$}}
\Text(-215,105)[c]{\scalebox{1.6}[1.6]{$g\gtrsim 0$}} 
\Text(-215,45)[c]{\scalebox{1.6}[1.6]{$g>g_{\text{cr}}$}}
\Text(-90,165)[c]{\scalebox{1.6}[1.6]{$T>0,g=0$}} 
\end{picture}
}
\end{center}
\medskip
\caption{Sketch of a typical $\beta$ function for the fermionic
  self-interactions $\lambda_i$ (taken from \cite{Braun:2006jd}): at zero gauge coupling, $g=0$ (upper
  solid curve), the Gau\ss ian fixed point $\lambda_i=0$ is IR
  attractive. For small $g\gtrsim 0$ (middle/blue solid curve), the
  fixed-point positions are shifted by the gauge-field fluctuations $\sim g^4$. For gauge
  couplings larger than the critical coupling $g>g_{\text{cr}}$
  (lower/green solid curve), no fixed points remain and the
  self-interactions quickly grow large, signaling \xsb. For increasing
  temperature, the parabolas become broader and higher, owing to
  thermal fermion masses; this is indicated by the dashed/red 
  line.} 
\label{fig:parab}
\end{figure}

Within this truncation, a simple picture for the chiral dynamics arises, see
Fig.~\ref{fig:parab}: at weak gauge coupling, the RG flow generates quark
self-interactions of order $\lambda\sim g^4$ via the last term in
Eq.~\eqref{eqlambda} with a negligible back-reaction on the gluonic RG
flow. If the gauge coupling in the IR remains smaller than a critical value
$g<g_{\text{cr}}$, the $\lambda$ self-interactions remain bounded, approaching
fixed points $\lambda_\ast$ in the IR.  The fixed points correspond to a
shifted Gau\ss ian fixed point $\lambda_\ast^{\text{Gau\ss}}|_{g^2=0}=0$. At these fixed
points, the fermionic subsystem remains in the chirally invariant phase which
is indeed realized at high temperatures $T>T_{\rm cr}$.

The evaluation of the QCD RG flow in a covariant derivative expansion
\cite{Reuter:1997gx,Gies:2002af} includes further gauge-field 
operators as well as kinetic terms for the fermion. From the gauge dynamics, the 
running gauge coupling can be extracted including its dependence on temperature 
$T$ as well as flavor and color numbers.  Following Ref.~\cite{Braun:2005uj,Braun:2006jd}, 
the increase of the running coupling in the IR is weakened on average for both larger $T$ 
and larger $\Nf$, in agreement with general expectations.  In addition, also $g_{\text{cr}}$
depends on $T$ and $\Nf$, even though the $\Nf$ dependence is rather weak. For instance, 
the (non-universal) zero-temperature value of the critical coupling for an optimized RG scheme is
$\alpha_{\text{cr}}=g^2_{\text{cr}}/(4\pi)\simeq 0.8$ for $\Nc=3$ and a wide range of 
$\Nf$~\cite{Gies:2005as}. 

The $T$ dependence of $g_{\text{cr}}$ arises from the quark modes acquiring
thermal masses. This leads to a quark decoupling, requiring stronger
interactions for critical quark dynamics. In the $\beta$ function picture of
Fig.~\ref{fig:parab}, the $\lambda_i$ parabolas become broader with a
higher maximum; hence, the annihilation of the Gau\ss ian fixed point by
pushing the parabola below the $\lambda_i$ axis requires a larger~$g_{\text{cr}}$.

%
\FIGURE[t]{
\includegraphics[%
  clip,
  scale=0.8]{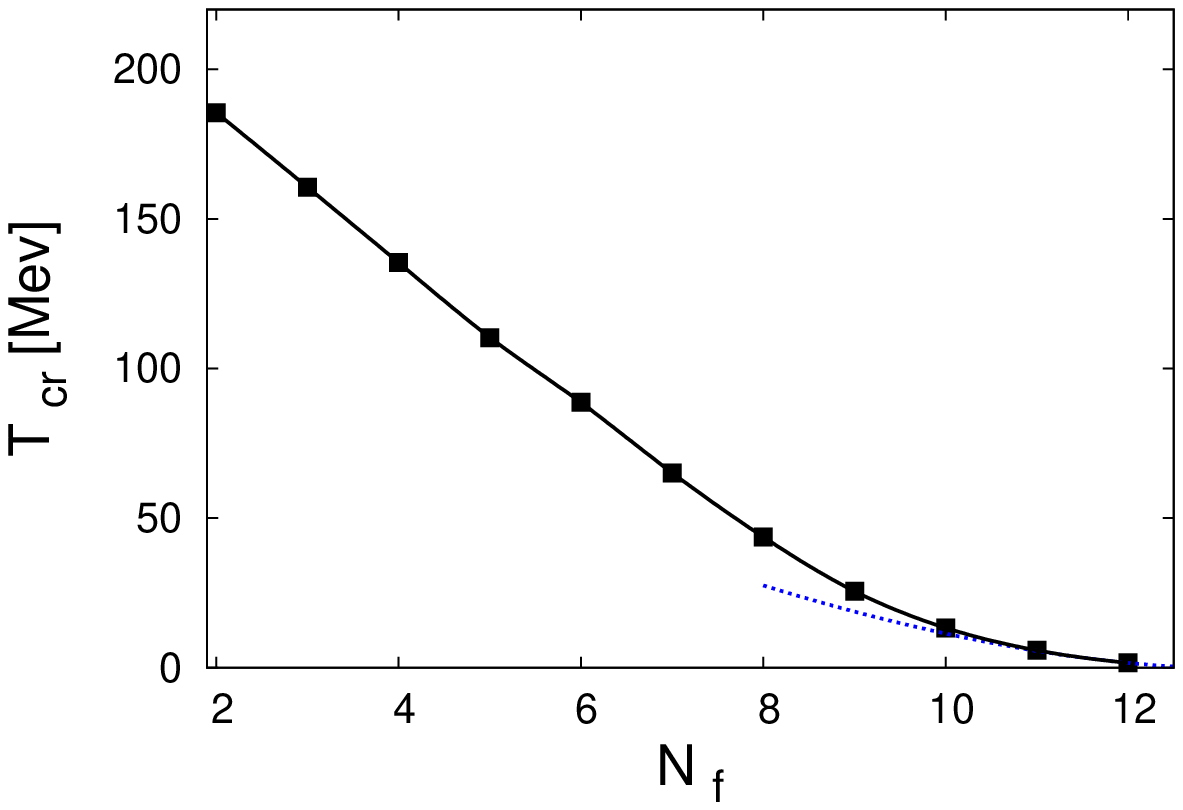}
\caption{Chiral-phase-transition temperature $T_{\text{cr}}$ versus
the number of massless quark flavors $\Nf$ for $\Nf\geq2$, as obtained
in Ref.~\cite{Braun:2006jd}. 
The flattening at $\Nf\gtrsim10$ is a consequence of the IR fixed-point
structure. The dotted line depicts the analytic estimate near
$\Nf^{\text{cr}}$ which follows from the fixed-point scenario (cf. Eq.~\eqref{eq:TcrTheta}).
}
\label{tc_nf} 
}
%

At zero temperature and for small $\Nf$, the IR fixed point $g_\ast^2$ is far
larger than $g^2_{\text{cr}}$, hence QCD is in the \xsb\ phase. For increasing
$T$, the temperature dependence of the coupling and that of $g^2_{\text{cr}}$
compete with each other.  For the case of many massless quark flavors $\Nf$,
the critical temperature is plotted in Fig.~\ref{tc_nf}. For the scale fixing
at the $\tau$ mass scale discussed above, we observe an almost linear decrease
of the critical temperature for increasing $\Nf$ with a slope of $\Delta
T_{\mathrm{cr}}=T(\Nf)-T(\Nf+1)\approx 25\,\mathrm{MeV}$ at small $\Nf$.  This linear
dependence of the full result confirms the simple estimate given in
\Eqref{eq:lowNfTc} to a very good accuracy. Moreover, the predicted relative
difference for $\Tc$ for $\Nf=2$ and $3$ flavors of $\Delta\simeq 0.146$ is in
very good agreement also with lattice studies \cite{Karsch:2000kv}. We
conclude that the shape of the phase boundary for small $\Nf$ is basically
dominated by fermionic screening.

For larger flavor number, the critical temperature decreases and the phase
transition line terminates at the zero-temperature quantum phase transition at
$\Nf^{\text{cr}}$, denoting the onset of the conformal window. In the RG study
of \cite{Braun:2005uj,Braun:2006jd}, we find a critical number of quark
flavors, $\Nf ^{\mathrm{cr}}\simeq12.9$.  This result for $\Nf^{\mathrm{cr}}$
agrees with other studies based on the 2-loop $\beta$ function
\cite{Banks:1981nn}. However, the precise value of $\Nf^{\mathrm{cr}}$ has to
be taken with care: for instance, in a perturbative framework,
$\Nf^{\mathrm{cr}}$ turns out to be sensitive to the 3-loop coefficient which
is not reliably reproduced in this leading-order study. This coefficient can
bring $\Nf^{\text{cr}}$ down to $\Nf^{\text{cr}}\simeq 10.0
\genfrac{}{}{0pt}{}{+1.6}{-0.7}$ which remains stable under the inclusion of
4-loop corrections~\cite{Gies:2005as}. The error bars parameterize truncation
errors which habe been quantified by artificial scheme dependencies. 

To the left of the conformal window ($\Nf < \Nf^{\text{cr}}$), 
the phase transition line shows a characteristic flattening. This is 
again in perfect agreement with our scaling relation \eqref{eq:TcrTheta}. 
The fit to numerical results from a functional RG approach is depicted by the dotted 
line in Fig.~\ref{tc_nf}. In particular, the fact that $|\Theta|<1$ 
near $\Nf^{\text{cr}}$ explains the flattening of the phase boundary near the 
critical flavor number. Within the covariant derivative expansion of RG 
flow, the IR critical exponent of the $\beta$ function of the 
coupling at the critical flavor number yields $\Theta(\Nf^{\text{cr}})\simeq -0.60$ 
\cite{Braun:2006jd}. However, this estimate is likely to be affected by truncation 
errors as $\Theta(\Nf)$ is expected to show sizable dependencies on $\Nf$; 
hence, any error in $\Nf^{\text{cr}}$ translates into a corresponding error in
$\Theta(\Nf^{\text{cr}})$.

%
\FIGURE[t]{
\includegraphics[%
  clip,
  scale=0.85]{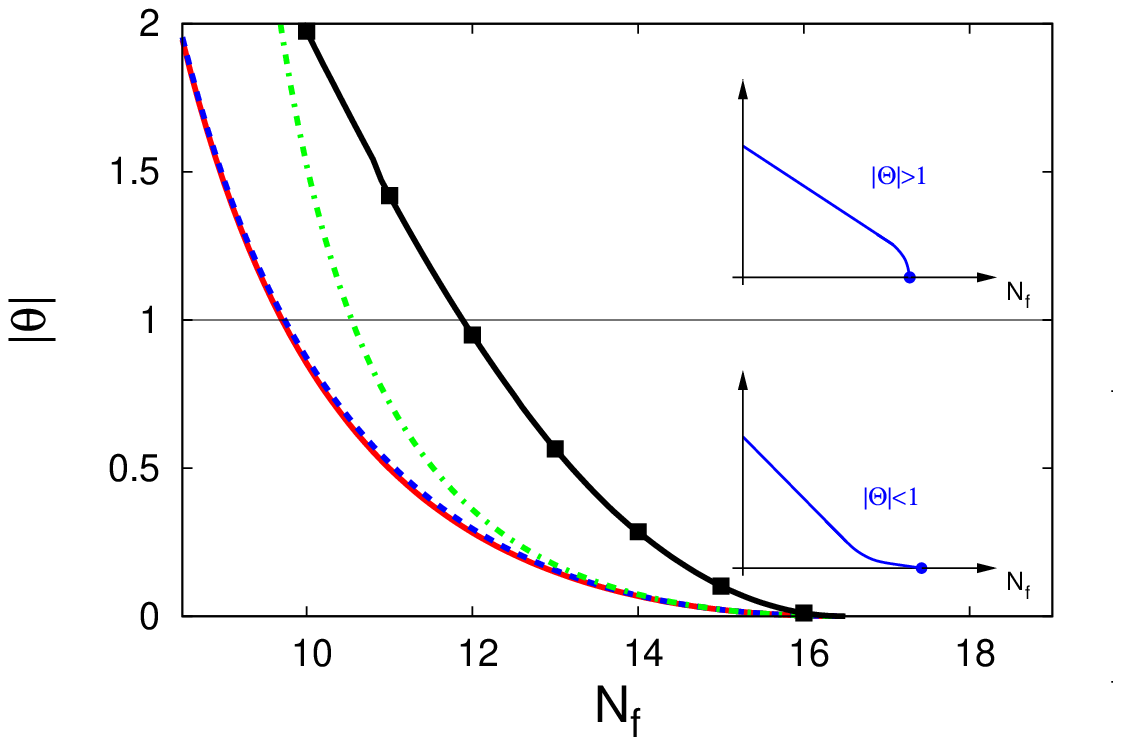} 
\caption{Critical exponent $|\Theta|$ of the running gauge coupling at the
  Caswell-Banks-Zaks fixed point as a function of the flavor number $\Nf$. The
  perturbative expansion appears to converge, as the difference between the
  3-loop (blue/dashed line) and 4-loop result (red/solid line) is below the
  1\% level (for $8.5< \Nf< 16.5$). The 2-loop result (green/dot-dashed line)
  shows larger deviations for smaller $\Nf$. The black/solid line with symbols
  corresponds
  to the estimate from the covariant derivative expansion of the RG flow
  \cite{Braun:2006jd}. The inlays depict the characteristic shapes of the
  scaling of a generic IR observable $\mathcal O$ for $|\Theta|>1$ (infinite
  slope at $\Nf^{\text{cr}}$) and $|\Theta|<1$ (vanishing slope at
  $\Nf^{\text{cr}}$). }
\label{fig:perttheta}
}
%

For comparison, we plot the perturbative estimates for the critical exponent
$\Theta(\Nf)$ in Fig.~\ref{fig:perttheta} based on the 4-loop $\beta$ function
in the $\overline{\text{MS}}$ scheme \cite{vanRitbergen:1997va}. The apparent
convergence of the perturbative expansion is remarkable as the difference
between the 3- and 4-loop result is below the 1\% level (for $8.5< \Nf<
16.5$). The 2-loop result shows larger deviations for smaller $\Nf$, as the
fixed-point coupling $g_\ast^2$ is larger in this regime. In
Fig.~\ref{fig:perttheta}, we also show the estimate of $\Theta$ from the
covariant derivative expansion of the RG flow \cite{Braun:2006jd}. While this
estimate includes nonperturbative contributions to all-loop orders in the
gauge sector, the derivative expansion in the fermion sector effectively
corresponds to the inclusion of just the (RG-improved) one-loop quark
diagram. This explains a large part of the difference to the perturbative
estimates. {Incidentally, results from further studies of the full
  $\beta$ function can be used to estimate $\Theta$. For instance, the
  $\Theta$ values obtained from the conjectured ``NSVZ-inspired'' $\beta$
  function \cite{Ryttov:2007cx} are identical to the 2-loop result.}

Qualitatively, the scaling of physical observables near the conformal window
shows a characteristic difference for $|\Theta|$ being larger or smaller than
one. If $|\Theta|>1$ the value of a given observable as a function of the
distance to the conformal window, $\mathcal O=\mathcal
O(\Nf^{\text{cr}}-\Nf)$, approaches the $\Nf$ axis with infinite slope. For
$|\Theta|<1$ the slope vanishes on the quantum critical point
$\Nf=\Nf^{\text{cr}}$. This characteristic shape dependence of the scaling is
again a firm prediction of our scenario and may be observable in lattice
simulations even away from the critical flavor number. 

Let us come back to the issue of choosing a specific fixing scale for
comparing theories with different $\Nf$ to each other. As stressed above,
e.g., the result for the shape of the phase boundary in Fig.~\ref{tc_nf} does
depend on our choice of fixing the running coupling at the $\tau$-mass scale
$m_\tau$. This choice is not unique: in principle, the fixing scale can be
chosen as a free function of $\Nf$. This would correspond to choosing an
arbitrary function $k_0=k_0(\Nf)$ for the global scale occurring in the
scaling relations~\eqref{eq:TcrTheta} and~\eqref{eq:observ_scaling}. Of
course, such a function induced by some ad-hoc scale fixing procedure could
obscure our scaling relations. For instance, an extreme choice would be given 
by measuring all dimensionful scales in units of the critical temperature
$T_{\text{cr}}$. In this case, the shape of the phase boundary would be a
horizontal line at $T/T_{\text{cr}}=1$ terminating at
$\Nf=\Nf^{\text{cr}}$. Nevertheless, the scaling relations could still be
verified, as they would translate into scaling relations for other external
scales: e.g., the scale $k$ at which the running coupling acquires a specific
value (say $\alpha=0.322$) would diverge with $\Nf\to\Nf^{\text{cr}}$
according to $k\sim T_{\text{cr}}|\Nf -\Nf^{\text{cr}}|^{\frac{1}{\Theta}}$
for fixed $T_{\text{cr}}$ (and $\Theta < 0$). This point of view can
constitute a different way of verifying our scaling relations on the
lattice. In summary, these considerations demonstrate that the existence of
scaling relations has a universal meaning, even though their concrete
manifestation can depend on the details of the $\Nf$-dependent scale
fixing. In particular, the non-analytic structure governed by the exponent
$\Theta$ always remains.
\section{Scaling in low-energy models}

Let us study explicitly the scaling of two model observables, namely the quark
condensate $\langle \bar{q}q\rangle$ and the (constituent) quark mass $M_q$,
by means of a simple ansatz for the effective potential $U$ for the low-energy
sector of QCD:
\begin{equation}
U(\phi) = \frac{1}{2}\text{Tr}\ln \left(\partial ^2 \!+\! M_{\sigma}^2 (\phi^2)\right)
 \!+\! \frac{N_f^2 \!-\!1}{2}\text{Tr}\ln \left(\partial ^2\! +\! M_{\pi}^2 (\phi^2)\right)
\! -\! N_f N_c \text{Tr}\ln \left(\fslash{\partial}\! +\! M_q (\phi)\right)\,.
\label{eq:eff_pot}
\end{equation}
Here, $\phi$ represents a bi-fermionic scalar mean-field, the expectation
value of which is related to the chiral condensate. Excitations on top of
this condensate correspond to the Goldstone bosons, say the pions, and the
sigma meson (radial mode). Mean-field expressions of this type generically
arise in many low-energy QCD models such as NJL-type models or the quark-meson
model. The $\Tr \ln$ terms simply correspond to the fluctuation contributions 
of the chiral mesonic and quark degrees of freedom. The masses $M_\pi$ and $M_\sigma$ of
these mesons depend on $\partial U/\partial \phi^2$ and a linear combination
of $\partial U/\partial \phi^2$ and $\partial^2 U/\partial \phi^2\partial
\phi^2$, respectively, see e.~g.~Refs.~\cite{Jungnickel:1995fp,Schaefer:1999em,Braun:2004yk}. 
The (constituent) quark mass $M_q$ is given by the product of the Yukawa coupling $h$
and $\phi$. 

The expression Eq.~\eqref{eq:eff_pot} for the effective potential
is UV divergent and needs to be regularized in some scheme (belonging to the
definition of the model) at an effective regulator scale\footnote{Additionally, 
IR divergences can occur in the broken phase in the chiral limit, as 
the Goldstone bosons are massless. Such IR divergencies can be tamed, 
e.g., by the functional RG, yielding well-defined IR predictions.}
$\Lambda_{\text{eff}}$.  The IR observables can then be computed by
solving Eq.~\eqref{eq:eff_pot} self-consistently for $U(\phi)$ together with
an initial renormalization condition provided at the scale
$\Lambda_{\text{eff}}$, e.g., $U(\phi)|_{\Lambda_{\text{eff}}}=\frac{1}{2}
m_{\Lambda_{\text{eff}}}^2 \phi^2+ \dots$. The bosonic mass parameter is
naturally related to the effective scale, $m_{\Lambda_{\text{eff}}}\sim
\Lambda_{\text{eff}}$. 

For simplicity, let us consider the model in the limit of large number of colors,
$N_c\to\infty$, for a fixed number of flavors $N_f$ by way of example. In this
limit only the quark loop survives and we are left with the following equation
for the effective potential $U$:
\begin{eqnarray}
U(\phi) &=& - N_f N_c \text{Tr}\ln \left(\fslash{\partial} + M_q (\phi)\right).
\label{eq:eff_pot2}
\end{eqnarray}
Note that the Yukawa coupling is constant in this limit and can therefore be
absorbed in a redefinition of the field $\phi$, i.e., $M_q = \phi$. The
resulting equation for the effective potential can now be solved easily. The
pion decay constant is given by the value of $\phi$ which minimizes the
potential $U$:
\be
\frac{\partial U}{\partial \phi}\Big|_{\phi=\phi_0}=0.
\ee
We find
\be
\phi_0 \simeq \sqrt{N_f N_c} \Lambda_{\text{eff}} = \sqrt{N_f N_c} k_0  |\Nf
-\Nf^{\text{cr}}|^{-\frac{1}{\Theta}}, 
\ee
where the last step holds near the conformal window, using the
relation \eqref{eq:kcrest2}.  Since the (constituent) quark mass is given by
$\phi_0$, we conclude that $M_q$ has a scaling behavior near $N_f
^{\text{cr}}$ identical to that of the critical temperature. We would like to
stress that the prefactor $\sqrt{N_f N_c}$ is an outcome of our large-$N_c$
analysis of the low-energy sector. In general, we expect that any observable
$\mathcal O$ comes along with a complicated pre-factor function $f_{\mathcal
  O}$ depending on the number of flavors $N_f$ and $N_c$. The 
determination of this function, e.~g. for the constituent mass, may become
complicated, depending on the truncations made in the low-energy
sector. However, we stress that the $N_f$ dependence coming from the prefactor
is subleading compared to the scaling with $|\Nf -\Nf^{\text{cr}}|$ according
to the IR critical exponent $|\Theta|$ of the running coupling.

The scaling of the quark condensate can then be obtained by employing the
following relation:
\be
|\langle \bar{q}q\rangle|=\frac{m_{\Lambda_\text{eff}}^2}{h^2}\phi_0\,.
\ee
Thus we find
\be
|\langle \bar{q}q\rangle| \sim k_0^3  |\Nf -\Nf^{\text{cr}}|^{-\frac{3}{\Theta}}\,.
\ee
The precise proportionality factor will again depend on the number of flavors
and colors which, however, cannot modify the scaling behavior with respect to
the distance to the conformal window. 

In \cite{Gromenko:2008tn}, the quark mass, the chiral condensate and the pion
decay constant have been computed within truncated Dyson-Schwinger equation
for many-flavor QCD. Signatures of the quantum critical point have been
identified and the critical exponents have been extracted from a fit to the
numerical data available away from the critical point for
$\Nf<\Nf^{\text{cr}}$. In the light of our scaling relation, the results of
\cite{Gromenko:2008tn} unfortunately remain somewhat inconclusive as
$\Nf^{\text{cr}}$ has been fitted for each IR observable separately (yielding
different values). This uncertainty is likely to spoil the fit for the
critical exponents. We expect that a more careful analysis in the vicinity of
the quantum critical point can easily put our scaling relation to test. 

Away from the chiral limit, the current quark mass is expected to modify the
scaling relations. A generalized Gell-Mann-Oakes-Renner relation based on the
fixed-point scenario in many-flavor QCD has been advocated in
\cite{Sannino:2008pz}. 

\section{Conclusions}

Many-flavor deformations of real QCD are a fascinating testing ground for
aspects of the chiral structure of QCD-like theories. The existence of a
conformal window for larger flavor numbers $\Nf$ (below the critical value where
asymptotic freedom is lost) gives rise to an interesting quantum critical
point on the $\Nf$ axis. Whereas a lot of effort has recently gone into the
determination of the value of $\Nf$, we have concentrated in this work on the
physics in the quantum critical region. Similarly to itinerant fermion
systems~\cite{Jakubczyk:2008cv}, we expect that the quantum critical point influences the properties
of the phase diagram of a number of observables as a function of $\Nf$ in the
neighborhood of the critical point. Note that the quantum phase transition in $\Nf$ can 
also be viewed as a higher-dimensional analogue of the Berezinskii-Kosterlitz-Thouless
transition~\cite{Kaplan:2009kr}. Similar transitions even occur in quantum
mechanics which can also be analyzed in an RG language \cite{Moroz:2009nm}.

We have shown in this work, that the scaling of generic IR observables in the
chirally broken phase as a function of $\Nf^{\text{cr}}-\Nf$ exhibits a
remarkably universal behavior. Our arguments were based on only few
assumptions: the existence of an IR Caswell-Banks-Zaks fixed point in the
running of the gauge coupling in the conformal window (which holds by
construction if the conformal window exists), and the existence of a critical
value of the gauge coupling for triggering chiral symmetry breaking. Whereas
the first assumption is a universal statement, the latter assumption needs to
be fulfilled only in specific RG schemes and for certain definitions of the
coupling. In other schemes and coupling definitions, this assumption may
translate into an assumption, for instance, on a sufficiently strong critical
behavior of the quark-gluon vertex.

We find that generic IR observables in the \xsb\ phase scale with the distance
to the conformal window in a characteristic fashion which is governed by only
one independent critical exponent. This critical exponent is directly related
to the corresponding critical exponent of the gauge coupling at the
Caswell-Banks-Zaks fixed point.  This relates universal quantities with each
other: chiral IR observables and the IR critical exponent. It establishes a
quantitative connection between the chiral structure ($T_{\text{cr}}$, $\langle
\bar\psi \psi\rangle$, $f_\pi$, etc.) and the IR gauge dynamics quantified by the
critical exponent $\Theta$. Most importantly, it is a parameter free
prediction following essentially from scaling arguments.

Recently, the interest in the conformal window of many-flavor QCD has been
revived by technicolor scenarios \cite{Weinberg:1979bn,Holdom:1981rm} for the
Higgs sector of the standard model which have attracted renewed attention
\cite{Hong:2004td}. As these scenarios are largely motivated by the
hierarchy problem of the standard model, it is important for some so-called
walking models that the technicolor sector remains in the vicinity of the
quantum critical point over a wide range of scales. The scaling laws should
therefore directly apply to such models, even though each model, of course,
has a fixed $\Nf$.

It should be stressed that varying $\Nf$ as a control parameter for a quantum
phase transition corresponds to comparing different theories with each
other. Such a comparison is not unique for non-conformal theories but requires
a specific choice of a dimensionful scale which is used as one and the same
ruler for different theories. In the present work, we have argued that
choosing a mid-momentum scale such as the $\tau$ mass has many advantages. For
small $\Nf$, this choice has lead us to a simple estimate for the $\Nf$
dependence of the critical temperature, which is in very good agreement with
lattice simulations. This type of argument has already been successfully
applied to the PNJL/PQM model, where implementing the linear $\Nf$ scaling for
small $\Nf$ has helped adjusting the physical parameters, leading to a
significant improvement of the thermodynamics properties of the model
\cite{Schaefer:2007pw}. 

Coming back to the scaling relation near the conformal window, we are aware of
the fact that, for instance, the relation \eqref{eq:TcrTheta} is difficult to
test by lattice gauge theory: neither the fixed-point scenario in the deep IR
nor large flavor numbers in the chiral limit are easily accessible.  However, given the
conceptual simplicity of the fixed-point scenario in combination with \xsb,
further lattice studies in this direction are certainly worthwhile.

\acknowledgments The authors thank J.M.~Pawlowski for useful discussions and
acknowledge support by the DFG under grants Gi~328/1-4, Gi~328/5-1 (Heisenberg
program) and FOR 723.

\end{document}